\documentclass{emulateapj}
\shorttitle{ENERGY VARYING SHOCK WAVE}
\shortauthors{SUZUKI \& SHIGEYAMA}
\begin{document}
\title{SELF-SIMILAR SOLUTIONS FOR THE EMERGENCE OF ENERGY VARYING SHOCK WAVES FROM PLANE-PARALLEL ATMOSPHERES}
\author{AKIHIRO SUZUKI\altaffilmark{1} and TOSHIKAZU SHIGEYAMA\altaffilmark{2}}
\altaffiltext{1}{Department of Astronomy, School of Science, University of Tokyo, Bunkyo-ku, Tokyo 113-0033, Japan.}
\altaffiltext{2}{Research Center for the Early Universe, Graduate School of Science, University of Tokyo, Bunkyo-ku, Tokyo 113-0033, Japan.}
\begin{abstract}
We present a self-similar solution to describe the propagation of a shock wave 
whose energy is deposited or lost at the front. 
Both of the propagation of the shock wave in a medium 
having a power-law density profile and the expansion of the medium to a vacuum 
after the shock breakout are described with a Lagrangian coordinate. 
The Chapman-Jouguet detonation is found to accelerate the medium most effectively. 
The results are compared with some numerical simulations in the literature. 
We derive the fractions of the deposited/lost energy at the shock front 
in some specific cases, which will be useful 
when applying this solution to actual phenomena.
\end{abstract}
\keywords{hydrodynamics --- shock waves --- supernovae: general}

\section{INTRODUCTION}
The self-similar solution is a powerful method to describe flows of a fluid involving strong shock waves.
In a supernova (SN) explosion, a shock wave propagates in the stellar interior toward the surface. 
This process can be regarded as a self-similar problem as long as the density distribution has a power-law profile with respect to the distance from the center or the surface.

In fact, self-similar solutions for this problem have been discovered and widely known 
\citep[see, e.g.,][]{zel02}. \citet{sak60} discovered a self-similar solution for the emergence of a strong shock wave from a medium with a power-law density distribution, which was first posed by \citet{gf56}. This solution describes distributions of the velocity, pressure and density of the matter with an Eulerian coordinate until the shock wave reaches the surface. After that, the solution uses a Lagrangian coordinate. \citet{mm99} invented another description for a whole evolution of this solution with a Lagrangian coordinate.

However, these solutions do not completely reflect on the circumstances.
In reality, thermal energy can be deposited and/or lost across a shock front 
(referred to as the energy variation in this paper). 
For example, some nuclear reactions can take place in the region behind the shock front when the temperature thereof becomes sufficiently high. This happens in a detonation wave and might be realized in a type Ia supernova (SN Ia), which is thought to be a thermonuclear explosion of a white dwarf. 
The thermonuclear explosion model for SNe Ia was originally proposed 
by \citet{hf60}, 
and several models for the propagation of the flame front have been discussed, 
for example, ``deflagration model''\citep[see][]{i74,n76}, 
''delayed detonation model''\citep[see][]{k91,y92,ww94}, and 
``pulsating delayed detonation model"\citep[][]{k91b}. 
In particular, the (pulsating) delayed detonation models suggest that the detonation wave is generated somewhere in the outer layer of a white dwarf and propagates toward the surface. 
In addition, a carbon detonation model of a white dwarf with the mass of $\sim 1\,M_\odot$ was proposed \citep{Shigeyama92} for a peculiar SN Ia, SN 1990N. The detonation wave propagates toward the surface also in this model. 
Furthermore, thermal energy generated behind the shock front can be transported through the front near the stellar surface where the optical depth is close to unity. In this region, photons diffuse out at velocities close to the speed of light, which is larger than the shock velocity. 
In other words, the energy of the shock wave 
can decrease by the radiative cooling.
\citet{ch76} calculated the effect of the cooling 
using a radiation-hydrodynamics code. 

This kind of phenomena in some type I SNe accelerates a considerable amount of matter to sufficiently high energies to synthesize light elements, Li, Be, and B through spallation reactions 
\citep[see][]{fields,ns04,n06}.
Thus the solution derived in this paper might be useful to investigate the chemical evolution of light elements in galaxies.

\citet{bs70} studied the treatment of the energy variation 
at the shock front in the frame work of self-similar solutions 
and showed that it can be treated by modifying the jump conditions 
at the front to impose the condition that the energy deposition or loss per unit mass 
be proportional to the internal energy. 
In real shock acceleration problem, this condition is rarely satisfied. 
Whether the energy is deposited by nuclear reactions or lost by radiation, it tends to have a characteristic energy per nucleon or a characteristic time scale. Thus the self-similarity of the associated flow might appear in very limited space and/or time if any.

 \begin{table*}[t]
\begin{center}
\caption{EIGEN VALUES $\lambda$ FOR $\gamma=4/3$,\label{eigen}}
\begin{tabular}{ccccccccccc}\tableline\tableline
&\multicolumn{9}{c}{$\gamma_\mathrm{e}$}\\\cline{2-11}
$\alpha$&11/3&3.0&2.0&1.5&4/3&1.2&1.1&1.01&1.001&1.0001\\\hline
3&1.0000&0.9525&0.7915&0.6407&0.5572&0.4608&0.3484&0.1212&0.03733&0.01092\\
1.5&0.5000&0.4725&0.4000&0.3271&0.2863&0.2386&0.1822&0.06507&0.02050&0.006113\\
\hline
\end{tabular}
\end{center}
\end{table*}

In the next section, we review the procedure of \citet{baren} to 
treat the energy variation. 
In $\S$ 3, we formulate the problem following \citet{mm99}. 
The integrations of the governing equations are found 
to be reduced to an eigen value problem. 
The Appendix gives a method to determine the eigen value. 
The results are presented in \S 4. 
Furthermore, the asymptotic behaviors of the velocity, pressure, and density of 
the ejecta are derived.
The energy distributions are deduced from the exponents of these physical quantities (\S 5). 
We conclude this paper in $\S$ 6.

\section{TREATMENT OF ENEGY VARIATION}
In this section, we review the procedure to treat the energy deposition or loss at the shock front introduced in \citet{baren}.
The shock propagates in the ideal gas whose density is $\rho_0$. 
We assume that the gas ahead of the shock front is stationary and neglect 
its pressure. 
The gas behind the shock front has the velocity $u_\mathrm{f}$, the pressure $p_\mathrm{f}$, 
and the density $\rho_\mathrm{f}$. 
Then the energy conservation law 
across the shock front propagating at the speed $U$ is described as 
\begin{equation}
\rho_\mathrm{f}(u_\mathrm{f}-U)\left[\frac{\gamma}{\gamma-1}\frac{p_\mathrm{f}}{\rho_\mathrm{f}}+
\frac{(u_\mathrm{f}-U)^2}{2}\right]=-\frac{\rho_0U^3}{2}.
\label{energy}
\end{equation}
Using the mass conservation law 
\begin{equation}
\rho_\mathrm{f}(u_\mathrm{f}-U)=-\rho_0U,
\end{equation}
and the momentum conservation law
\begin{equation}
\rho_\mathrm{f}(u_\mathrm{f}-U)u_\mathrm{f}+p_\mathrm{f}=0,
\end{equation}
we obtain
\begin{equation}
\rho_\mathrm{f}(u_\mathrm{f}-U)\left[\frac{p_\mathrm{f}}{(\gamma-1)\rho_\mathrm{f}}+\frac{u_\mathrm{f}^2}{2}-q\right]
+p_\mathrm{f}u_\mathrm{f}=0,
\label{energy2}
\end{equation}
where we have added an energy flux term $-\rho_\mathrm{f}(u_\mathrm{f}-U)q$ to the left hand side of equation (\ref{energy}). 
Here the energy source term is denoted by $q$. 
This source term needs to be proportional to the internal energy to guarantee the self-similarity of the problem. Thus the source term is expressed as 
\begin{equation}
q=\frac{\gamma_\mathrm{e}-\gamma}{(\gamma_\mathrm{e}-1)(\gamma-1)}\frac{p_\mathrm{f}}{\rho_\mathrm{f}},
\end{equation}
by introducing the effective adiabatic exponent $\gamma_\mathrm{e}$ \citep{bs70}. The physical meaning of $\gamma_\mathrm{e}$ is as follows;
(1) if $\gamma_\mathrm{e}>\gamma$,  energy is deposited at the front. 
(2) if $\gamma_\mathrm{e}=\gamma$, the situation is same as the classical very 
intense explosion problem. 
(3) if $\gamma_\mathrm{e}<\gamma$, energy is lost at the shock front. 
Using the effective adiabatic exponent $\gamma_\mathrm{e}$, 
we transform equation (\ref{energy2}) to
\begin{equation}
\rho_\mathrm{f}(u_\mathrm{f}-U)\left[\frac{p_\mathrm{f}}{(\gamma_\mathrm{e}-1)\rho_\mathrm{f}}+\frac{u_\mathrm{f}^2}{2}\right]
+p_\mathrm{f}u_\mathrm{f}=0.
\label{energy3}
\end{equation}
This expression has the same form as the classical very intense 
explosion problem except for replacing $\gamma$ with $\gamma_\mathrm{e}$.
Therefore, the Rankine-Hugoniot relations for a strong shock become
\begin{equation}
u_\mathrm{f}=\frac{2}{\gamma_\mathrm{e}+1}U,
\label{RH1}
\end{equation}
\begin{equation}
p_\mathrm{f}=\frac{2}{\gamma_\mathrm{e}+1}\rho_0U^2,
\label{RH2}
\end{equation}
\begin{equation}
\rho_\mathrm{f}=\frac{\gamma_\mathrm{e}+1}{\gamma_\mathrm{e}-1}\rho_0,
\label{RH3}
\end{equation}
In other words, we can treat the energy variation 
by varying $\gamma_\mathrm{e}$

\section{FORMULATION}
In this section, we derive equations describing the self-similar motion
of the stellar matter.
Following \citet{mm99}, 
we treat the shock emergence with a Lagrangian coordinate.
\subsection{Basic equations} 
We first define the space coordinate $x$ as the distance of the fluid element from the stellar surface 
and the time coordinate $t$ as the time measured from the moment when the shock reaches the surface.
The initial position of each fluid element is denoted by $x_0$.
The basic equations to describe the evolution of the density $\rho$, velocity $u$ and pressure $p$ of the fluid are formulated with the Lagrangian coordinate $x_0$ as
\begin{equation}
\frac{\partial x}{\partial x_0}=\frac{\rho_0}{\rho},
\label{base1}
\end{equation}
\begin{equation}
u=\frac{\partial x}{\partial t},
\label{base2}
\end{equation}
\begin{equation}
\frac{\partial u}{\partial t}=
-\frac{1}{\rho_0}\frac{\partial p}{\partial x_0},
\label{base3}
\end{equation}
\begin{equation}
p\rho^{-\gamma}=p_\mathrm{f}\rho_\mathrm{f}^{-\gamma}.
\label{base4}
\end{equation} 
We assume that the stellar envelope has a power-law density profile; 
\begin{equation}
\rho_0(x)=\left\{\begin{array}{cc}
k_1x_0^{\alpha}&{\rm for}\ x_0\geq 0,\\
0&{\rm for}\ x_0<0,
\end{array}\right.
\label{profile}
\end{equation}
where $k_1$ and $\alpha$ are constants.
If the envelope of the star is radiative, the power-law exponent is $\alpha=3$.
On the other hand, $\alpha=1.5$ corresponds to the convective envelope. 
The time when each fluid element experiences the shock is denoted by $t_0(<0)$. 
By defining the variable $m$ as
\begin{equation}
m=\int^{x_0}_{0}\rho_0(y)dy,
\end{equation}
the coordinates $x$, $x_0$, and $t_0$ are labeled by $m$.  
Then, we define the similarity variable $\eta$ as
\begin{equation}
\eta=\frac{t}{t_0(m)},
\label{eta}
\end{equation}
and the flow variable $S(\eta)$ as the position of a fluid element normalized with its initial position, i.e.,
\begin{equation} 
S(\eta)=\frac{x(m)}{x_0(m)}.
\label{flowv}
\end{equation}
Both of these variables $\eta(m)$ and $S(\eta)$ become unity when the shock hits the fluid element with $m$ whereas $\eta$ becomes zero when the shock reaches the surface. The phase with $0<\eta<1$ corresponds to the evolution of shocked fluid elements till the shock breakout and the phase with $\eta<0$ to the expansion of the stellar matter to a vacuum.

\begin{figure}
\includegraphics[scale=0.7,angle=270]{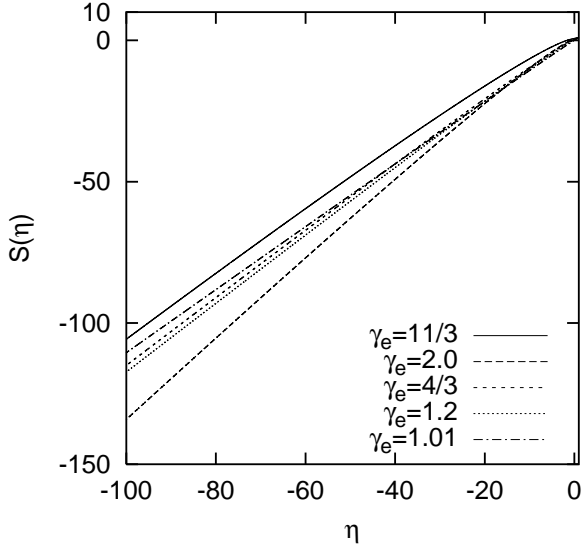}
\caption{Profile of the flow variable $S(\eta)$ for $\alpha=3.0, \gamma=4/3, \gamma_\mathrm{e}=11/3,2.0,4/3,1.2,1.01$ \label{coms}}
\end{figure}

\subsection{Expression of physical variables}
Next we derive relations between the flow variable $S(\eta)$ and 
the velocity, pressure and density of the stellar matter. 
We assume that the velocity of the shock front takes the form of some power of the original position $x_0$ 
of the fluid element;
\begin{equation}
U=-k_2[x_0(m)]^{-\lambda},
\label{shockv}
\end{equation}
where $k_2$ and $\lambda$ are constants. The latter will be determined as the eigen value. 
Then we obtain the following relation by integrating equation (\ref{shockv}) with respect to time from $t=0$ to $t=t_0$
\begin{equation}
t_0=-\frac{x_0^{1+\lambda}}{k_2(1+\lambda)}.
\end{equation}
On the other hand, a differentiation of both sides of equation (\ref{eta}) with respect to $x_0$ leads to
\begin{equation}
d\eta=-\eta(1+\lambda)\frac{dx_0}{x_0}
\label{deta}
\end{equation}
Using equations (\ref{flowv}),(\ref{deta}), and (\ref{base1}), the density is expressed as
\begin{equation}
\rho(\eta,x_0)=
\frac{\rho_0}{S(\eta)-(\lambda+1)\eta S'(\eta)},
\label{dens}
\end{equation}
where the prime denotes the derivative with respect to $\eta$.  
Then the pressure is also expressed in terms of $S(\eta)$ as
\begin{eqnarray}
p&=&p_\mathrm{f}\left(\frac{\rho}{\rho_\mathrm{f}}\right)^{\gamma}\nonumber\\
&=&\frac{2}{\gamma_\mathrm{e}+1}
\left(\frac{\gamma_\mathrm{e}-1}{\gamma_\mathrm{e}+1}\right)^{\gamma}\rho_0U^2
\label{pres}\\
&&\times\left[S(\eta)-(\lambda+1)\eta S'(\eta)\right]^{-\gamma},\nonumber
\end{eqnarray}
the velocity
\begin{equation}
u=(1+\lambda)S'U.
\label{vel}
\end{equation}
As a result, solving equations (\ref{base1}) through (\ref{base4}) 
reduces to the integration of a second order ordinary differential equation for the flow variable $S(\eta)$. 
The method to determine the function $S(\eta)$ is described in Appendix.

\section{RESULTS}
\begin{figure}
\includegraphics[scale=0.7,angle=270]{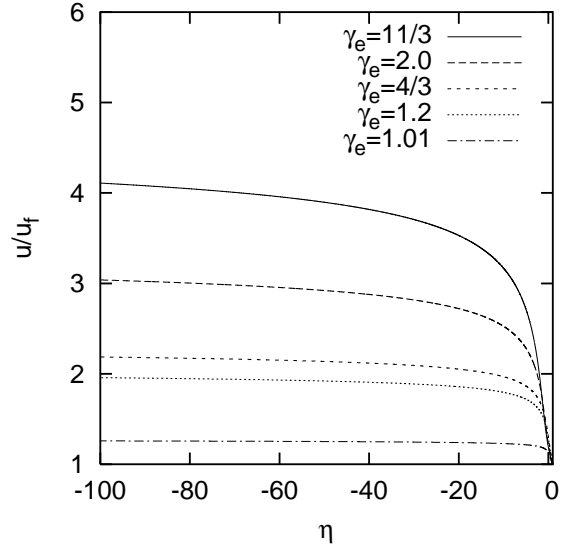}
\caption{Profile of the velocity $u/u_\mathrm{f}$ for $\alpha=3.0, \gamma=4/3, \gamma_\mathrm{e}=11/3,2.0,4/3,1.2,1.01$ \label{comv}}
\end{figure}
\begin{figure}
\includegraphics[scale=0.7,angle=270]{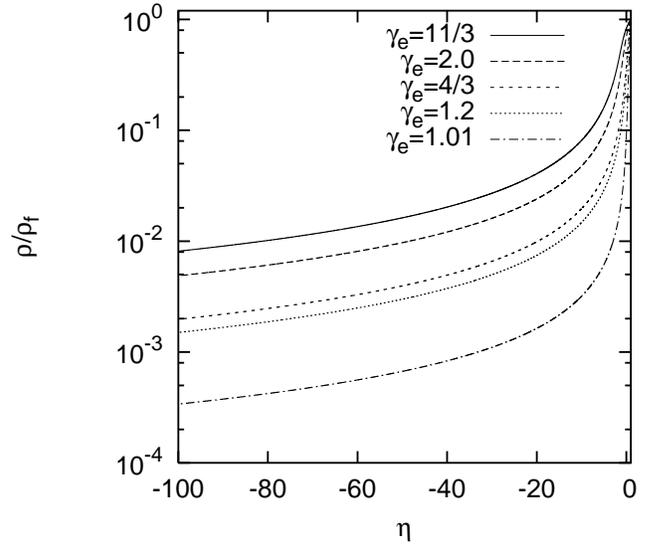}
\caption{Profile of the density $\rho/\rho_\mathrm{f}$ for $\alpha=3.0, \gamma=4/3, \gamma_\mathrm{e}=11/3,2.0,4/3,1.2,1.01$ \label{comr}}
\end{figure}

\subsection{Distributions of the physical variables} 
\begin{figure}
\includegraphics[scale=0.7,angle=270]{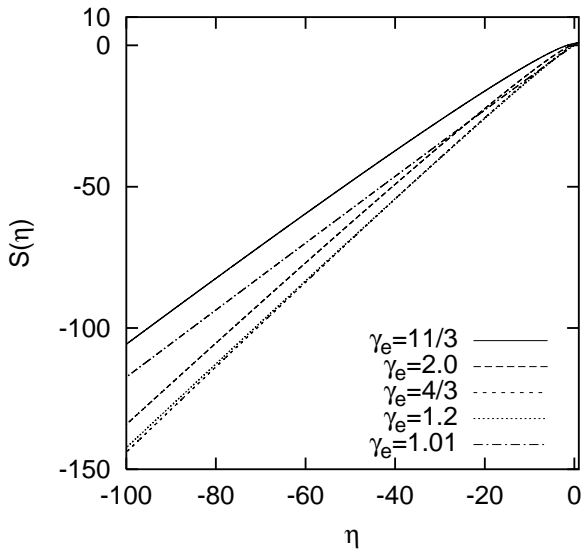}
\caption{Same as Fig. \ref{coms} but for $\alpha=1.5$ \label{comsr}}
\end{figure}

\begin{figure}
\includegraphics[scale=0.7,angle=270]{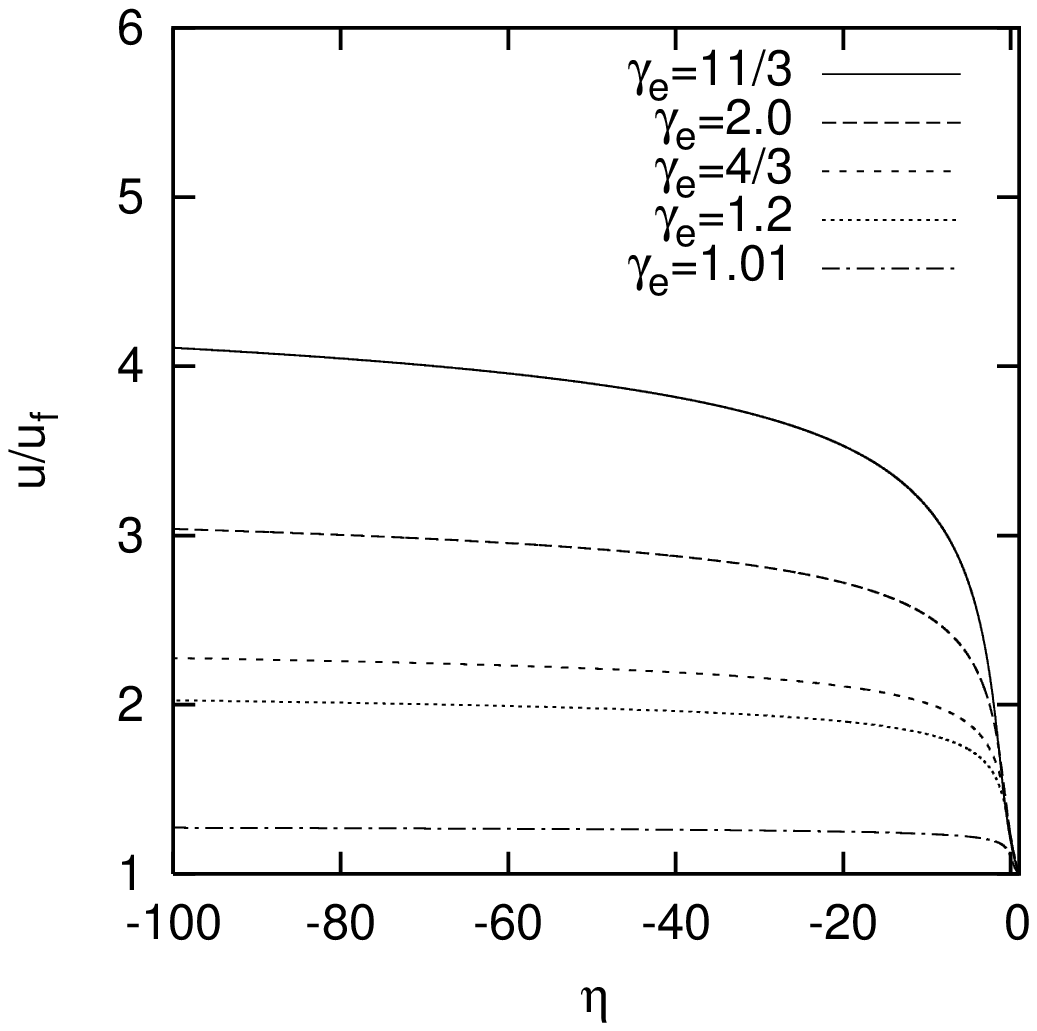}
\caption{Same as Fig. \ref{comv} but for $\alpha=1.5$ \label{comvr}}
\end{figure}

\begin{figure}
\includegraphics[scale=0.7,angle=270]{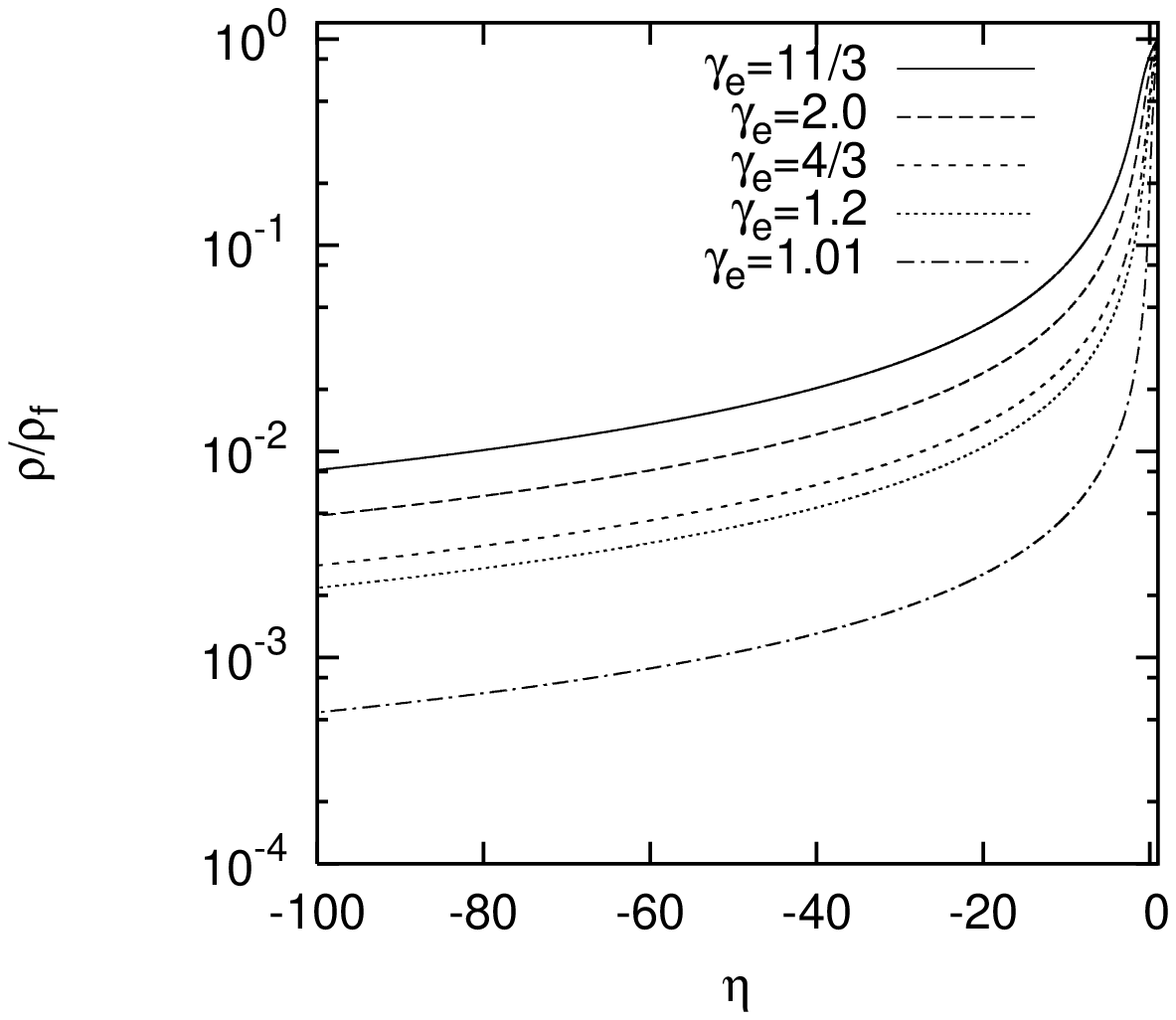}
\caption{Same as Fig. \ref{comr} but for $\alpha=1.5$ \label{comrr}}
\end{figure}

The distributions of the flow variable $S(\eta)$, 
velocity, and density of the stellar matter 
for $\alpha=3$ and various $\gamma_\mathrm{e}$'s are shown 
in Figures \ref{coms},\ref{comv}, and \ref{comr}, respectively.
The same distributions but for $\alpha=1.5$ are shown 
in Figures \ref{comsr},\ref{comvr}, and \ref{comrr}.
Each line in these figures traces the time evolution of each fluid element after the arrival of the shock front (at $\eta=1$).
At the same time, each line in the range of $0<\eta<1$ represents the spatial distribution of the shocked fluid element before the shock breakout while lines with $\eta<0$ represents the spatial distribution after the shock breakout.
The results show that the energy variation at immediately behind the shock front pronounces the acceleration after the shock breakout rather than before the shock breakout (Figs. \ref{comv}, \ref{comvr}). 
All the numerical results suggest that the velocities converge to constant values and that the density distributions converge to power-law profiles as $t\rightarrow \infty$. 
Such asymptotic behaviors are discussed in next section.

\subsection{Asymptotic behavior}
From the numerical results, the derivative of the flow variable $S'$ is found to converge to a constant value, 
thus the asymptotic behavior of $S$ should be
\begin{equation}
S\propto \eta.
\end{equation}
Using this relation, we derive the power-law exponents of velocity, pressure, and density 
of the ejecta with respect to the variable $x_0$ as
\begin{equation}
u\propto x_0^{-\lambda},\ \ \ p\propto x_0^{\alpha-2\lambda+(1+\lambda)\gamma},\ \ \ 
\rho\propto x_0^{\alpha+\lambda+1}.
\label{asym}
\end{equation}
The values of the exponents of $p,\rho$ for various $\gamma_\mathrm{e}$ are shown in Tables \ref{asym_p} and \ref{asym_r}.  
From these relations, we obtain
\begin{equation}
\rho\propto u^{-(\alpha+\lambda+1)/\lambda}.
\end{equation}
Thus the density distribution in the free expansion phase can be obtained by substituting $u=x/t$ into this expression.
The exponents in this equation are also shown in Table \ref{asym_u}.

The late time behavior for the case including the energy variation is essentially the same as adiabatic case ($\gamma_\mathrm{e}=\gamma$) except for the value of the exponents. 

\subsection{Chapmann-Jouguet detonation}
\begin{table*}[t]
\begin{center}
\caption{POWER-LAW EXPONENT OF $p$ FOR $\gamma=4/3$,\label{asym_p}}
\begin{tabular}{cccccccccccc}\tableline\tableline
&\multicolumn{9}{c}{$\gamma_\mathrm{e}$}\\\cline{3-12}
$\alpha$&$t$&11/3&3.0&2.0&1.5&4/3&1.2&1.1&1.01&1.001&1.0001\\\hline
3&0&1.000&1.095&1.417&1.719&1.886&2.078&2.303&2.758&2.925&2.978\\
&$+\infty$&3.667&3.698&3.806&3.906&3.962&4.026&4.101&4.253&4.308&4.326\\
1.5&0&0.5000&0.5550&0.7000&0.8459&0.9275&1.023&1.136&1.370&1.459&1.488\\
&$+\infty$&2.500&2.518&2.567&2.615&2.642&2.674&2.712&2.790&2.820&2.829\\
\tableline
\end{tabular}
\end{center}
\end{table*}
\begin{table*}[t]
\begin{center}
\caption{POWER-LAW EXPONENT OF $\rho$ FOR $\gamma=4/3$,\label{asym_r}}
\begin{tabular}{cccccccccccc}\tableline\tableline
&\multicolumn{9}{c}{$\gamma_\mathrm{e}$}\\\cline{3-12}
$\alpha$&$t$&11/3&3.0&2.0&1.5&4/3&1.2&1.1&1.01&1.001&1.0001\\\hline
3&0&3.000&3.000&3.000&3.000&3.000&3.000&3.000&3.000&3.000&3.000\\
&$+\infty$&5.000&4.953&4.792&4.641&4.557&4.461&4.348&4.121&4.037&4.011\\
1.5&0&1.500&1.500&1.500&1.500&1.500&1.500&1.500&1.500&1.500&1.500\\
&$+\infty$&3.000&2.973&2.900&2.827&2.786&2.739&2.682&2.565&2.521&2.506\\
\hline
\end{tabular}
\end{center}
\end{table*}

We concentrate on the flow called "Chapmann-Jouguet detonation" 
that satisfies the Chapmann-Jouguet condition;
the velocity of the gas relative to the shock front 
is equal to the local velocity of sound. 
\citet{bs70} showed that this condition can be expressed as the relation between $\gamma$ and $\gamma_\mathrm{e}$. 
We review the procedure and apply it to this study. 

From the Rankine-Hugoniot relations (\ref{RH1})-(\ref{RH3}), 
the velocity of the gas relative to the shock front is
\begin{equation}
|u_\mathrm{f}-U|=\frac{\gamma_\mathrm{e}-1}{\gamma_\mathrm{e}+1}U,
\label{cj1}
\end{equation}
the velocity of sound is
\begin{equation}
\sqrt{\frac{\gamma p_\mathrm{f}}{\rho_\mathrm{f}}}=
\frac{\sqrt{2\gamma(\gamma_\mathrm{e}-1)}}{\gamma_\mathrm{e}+1}U.
\label{cj2}
\end{equation} 
Equating the right hand sides of equations (\ref{cj1}) and (\ref{cj2}), 
the condition is expressed as 
\begin{equation}
\gamma_\mathrm{e}=2\gamma+1.
\end{equation}
Using this expression, $T_\mathrm{f}$ and $T_\mathrm{s}$ in (\ref{front}) and (\ref{sing}) 
are found to take the same value.
This means that the singular point is located at the shock front.
Then by equating $R$ defined in (\ref{varR}) at the shock front to that at the singular point, 
we obtain the analytic expression for $\lambda$ as
\begin{equation}
\lambda=\frac{\alpha}{3}.
\label{cj}
\end{equation}
On the other hand, the following relation is derived from equations (\ref{profile}) and (\ref{shockv});
\begin{equation}
U\propto\rho_0^{-\lambda/\alpha}.
\end{equation}
In C-J detonation, substitution of the relation (\ref{cj}) yields  
\begin{equation}
U\propto\rho_0^{-1/3},
\end{equation}
which is independent of $\alpha$.
Therefore this relation can be applied to an arbitrary power-law density profile. 
For a shock wave without energy variation ($\gamma_\mathrm{e}=\gamma$), substitution of the values in Table \ref{eigen} yields
\begin{equation}
U\propto\rho_0^{-0.19},
\end{equation}
which is also independent of $\alpha$.

\subsection{Energy loss limit}
\begin{figure}
\includegraphics[scale=0.7,angle=270]{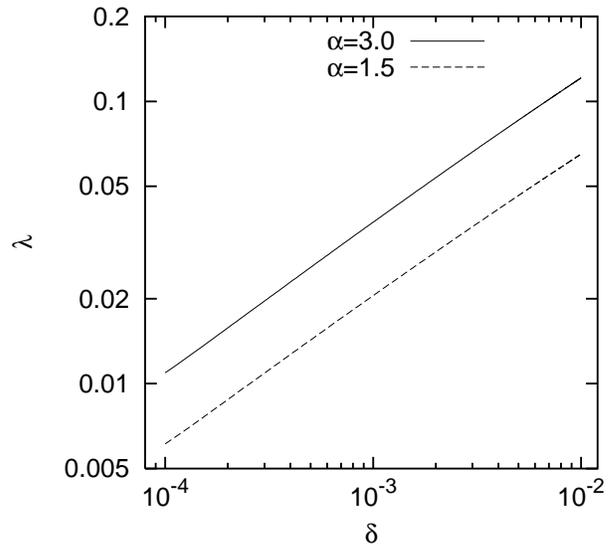}
\caption{The behavior of $\lambda$ as a function of $\delta$ for $\gamma=4/3, \alpha=3.0, 1.5$ \label{lim}}
\end{figure}

Here we consider the situation that $\gamma_\mathrm{e}\rightarrow 1$. 
However the density diverges for the case that $\gamma_\mathrm{e}$ is exactly unity. 
Therefore introducing the parameter $\delta$ as
\begin{equation}
\delta =\gamma_\mathrm{e}-1,  
\end{equation}
we treat the behavior of  the eigen value $\lambda$ when $\delta$ is close to zero. 
Fig. \ref{lim} shows the behavior in the range of $10^{-4}\leq\delta\leq 10^{-2}$ for $\gamma=4/3, \alpha=3.0, 1.5$. 
In this range, the following empirical relations are derived by interpolation;
 \begin{equation}
 \lambda=\left\{\begin{array}{cc}
1.3\delta^{0.51}&{\rm for}\ \alpha=3.0,\\
0.65\delta^{0.50}&{\rm for}\ \alpha=1.5.
\end{array}\right.
\end{equation}
These formulae suggest that when the shock becomes radiative, the shock velocity tends to be independent of the initial density. Thus the shock is no longer accelerated due to radiative energy loss and propagates at a constant velocity.
 
\section{ENERGY SPECTRUM}
In this section, we derive the energy spectrum from the asymptotic behavior 
of physical quantities described above. 
The energy spectrum is defined as the mass whose energy 
is larger than $\epsilon$;
\begin{equation}
M(>\epsilon)=\int^{\infty}_{\epsilon}\rho dx.
\end{equation}
From the relation (\ref{asym}) and $\epsilon=u^2/2$, 
the dependence on $\epsilon$ is found to be
\begin{equation}
M(>\epsilon)\propto \epsilon^{-(\alpha+1)/2\lambda}.
\end{equation}
The values of the power-law exponent for various $\gamma_\mathrm{e}$ are 
shown in Table \ref{spec}.  
Here $\gamma$ is assumed to be $4/3$. 
The results for $\gamma_\mathrm{e}=2\gamma+1$ give the upper limit for the exponent, 
because $\gamma_\mathrm{e}$ takes the maximum value for the Chapmann-Jouguet detonation and then $\lambda$ yields the maximum.
The exponents increase as the energy is deposited at the shock front and converge to the upper limit
in the Chapmann-Jouguet detonation. 
On the other hand, the energy loss decreases the exponents and 
the exponents diverge at $\gamma_\mathrm{e}=1$ related to the situation when 
all the internal energy is lost at the shock front.  
This result implies that the energy distribution of cosmic rays for the case 
$\gamma_\mathrm{e}>\gamma$ (energy deposition) 
becomes harder than that for the case $\gamma_\mathrm{e}=\gamma$.

\section{CONCLUSIONS AND DISCUSSION}
\begin{table*}[t]
\begin{center}
\caption{VELOCITY SPECTRUM FOR $\gamma=4/3$,\label{asym_u}}
\begin{tabular}{cccccccccccc}\tableline\tableline
&\multicolumn{9}{c}{$\gamma_\mathrm{e}$}\\\cline{3-12}
$\alpha$&$t$&11/3&3.0&2.0&1.5&4/3&1.2&1.1&1.01&1.001&1.0001\\\hline
3&0&-3.000&-3.150&-3.790&-4.683&-5.384&-6.511&-8.611&-24.76&-80.37&-274.8\\
&$+\infty$&-5.000&-5.199&-6.054&-7.244&-8.178&-9.681&-12.48&-34.01&-108.2&-367.4\\
1.5&0&-3.000&-3.174&-3.750&-4.586&-5.240&-6.288&-8.234&-23.05&-73.16&-245.4\\
&$+\infty$&-6.000&-6.291&-7.250&-8.644&-9.733&-11.48&-14.72&-39.42&-122.9&-409.9\\
\tableline
\end{tabular}
\end{center}
\end{table*}
\begin{table*}[t]
\begin{center}
\caption{ENERGY SPECTRUM FOR $\gamma=4/3$,\label{spec}}
\begin{tabular}{ccccccccccc}\tableline\tableline
&\multicolumn{9}{c}{$\gamma_\mathrm{e}$}\\\cline{2-11}
$\alpha$&11/3&3.0&2.0&1.5&4/3&1.2&1.1&1.01&1.001&1.0001\\\hline
3&-2.000&-2.100&-2.527&-3.122&-3.589&-4.340&-5.740&-16.50&-53.58&-183.2\\
1.5&-2.500&-2.645&-3.125&-3.822&-4.367&-5.141&-6.862&-19.21&-60.96&-204.5\\
\tableline
\end{tabular}
\end{center}
\end{table*}

In this paper, we have derived a self-similar solution to describe 
the propagation of the shock wave whose energy varies at the shock front. 
Using the solution, we describe the propagation of the shock wave 
in a plane-parallel medium having a power-law density profile. 
The motion of the ejecta after the shock breakout is derived. 
The asymptotic behavior of the ejecta is also derived 
and then the energy spectrum in the free expansion phase is deduced. 
Considering the case for $\gamma_\mathrm{e}=2\gamma+1$, 
we derive the power-law exponent of the energy spectrum analytically, 
which gives the upper limit for it. 

The calculation of the radioactive energy input model for type Ia SNe  
\citep[see][]{cm69} showed that the explosion of an $n=3$ polytrope  
leads to the outer density profile 
\begin{equation}
\rho\propto u^{-7},
\end{equation}
which corresponds to the case $\gamma_\mathrm{e}\approx 1.6$ in this work. 
Then, the total energy per unit mass at the shock front becomes
\begin{equation}
E_{\rm tot}=\left(\frac{1}{\gamma_\mathrm{e}-1}+\frac{1}{\gamma-1}\right)\frac{p_\mathrm{f}}{\rho_\mathrm{f}}
\approx \frac{14}{3}\frac{p_\mathrm{f}}{\rho_\mathrm{f}},
\end{equation}
and the energy source per unit mass becomes
\begin{equation}
q\approx \frac{4}{3}\frac{p_\mathrm{f}}{\rho_\mathrm{f}},
\end{equation}
which is $\sim 30$\% of the total energy at the front. 
For the Chapmann-Jouguet detonation, the outer density profile is
\begin{equation}
\rho\propto u^{-5},
\end{equation}
which corresponds to
\begin{equation}
E_{\rm tot}\approx \frac{27}{8}\frac{p_\mathrm{f}}{\rho_\mathrm{f}},
\end{equation}
and 
\begin{equation}
q\approx \frac{21}{8}\frac{p_\mathrm{f}}{\rho_\mathrm{f}},
\end{equation}
This implies that $\sim 78$\% of the total energy at the front is deposited by 
nuclear reactions. 

On the other hand, the shock wave in a SN II suffers 
from the radiative cooling. 
\citet{ch82} showed that the outer density profile of the ejecta of SNe II is 
steeper than that of SNe I, and adopted the relation
\begin{equation}
\rho\propto u^{-12}.
\label{u-12}
\end{equation}
We assume that the progenitor of the SN II is a red giant, 
whose stellar envelope is convective ($\alpha=1.5$). 
The relation (\ref{u-12}) corresponds to 
the case $\gamma_\mathrm{e}\approx 1.18$ in this work, 
which deduces to
\begin{equation}
E_{\rm tot}\approx \frac{77}{9}\frac{p_\mathrm{f}}{\rho_\mathrm{f}},
\end{equation}
and 
\begin{equation}
q\approx -\frac{23}{9}\frac{p_\mathrm{f}}{\rho_\mathrm{f}},
\end{equation}
then $\sim 30$\% of the total energy 
is lost at the shock front by the radiative cooling. 

Here we consider some applicable limits of this solution. 
First, from the comparisons with some numerical simulations, 
the treatment of the energy variation in this self-similar solution is found
to be valid. 
However, this solution can not determine the coefficients of physical variables 
because we cannot express $k_2$ in equation (\ref{shockv}) 
in terms of physical parameters specifying the phenomenon. 
As a consequence, 
we can not obtain absolute values for the physical variables. 
Second, we consider the behavior of the energy source $q$ near the surface, 
where the total energy per unit mass diverges 
because its density becomes zero. 
Simultaneously, the energy source diverges to $\pm\infty$ for the energy 
deposition/loss. 
We assume that the energy deposition is caused by some nuclear reactions. 
Then the energy generation rate must decrease near the surface, 
because the density decreases while the shock front accelerates toward the surface. 
Therefore, this model becomes unrealistic near the surface.
On the other hand, for the energy loss, 
the situation is realistic. 
As the total energy diverges, the energy source also diverges to $+\infty$.
This situation is understood as the radiative diffusion 
deprives the stellar matter of the diverging energy. 

\acknowledgments
This work has been partially supported by the grants in aid (16540213) of the Ministry of Education, Science, Culture, and Sports in Japan.

\appendix
\section{DERIVATION OF DISTRIBUTIONS}
\subsection{Formulation}
From the basic equations of gas dynamics (\ref{base2}) and (\ref{base3}), 
the equation of motion is expressed as
\begin{equation}
\frac{d^2x}{dt^2}=-\frac{1}{\rho}\left(\frac{dp}{dx}\right)_t
=-\frac{1}{\rho_0}\left(\frac{dp}{dx_0}\right)_t.
\label{eom}
\end{equation}
Using the flow variable $S(\eta)$, the second derivative of $x$ yields 
\begin{equation}
\frac{d^2x}{dt^2}=\frac{x_0}{t_0^2}S''(\eta).
\end{equation}
Substitution of the relation (\ref{pres}) into 
the right hand of equation (\ref{eom})  yields 
\begin{eqnarray}
\frac{1}{\rho_0}\left(\frac{dp}{dx_0}\right)_t&=&
\frac{2}{\gamma_\mathrm{e}+1}\left(\frac{\gamma_\mathrm{e}-1}{\gamma_\mathrm{e}+1}\right)^{\gamma}
k_2^2x_0^{-\alpha}
\frac{d}{dx_0}\frac{x_0^{\alpha-2\lambda}}
{[S(\eta)-(1+\lambda)\eta S'(\eta)]^{\gamma}}\nonumber\\
&=&\frac{2}{\gamma_\mathrm{e}+1}\left(\frac{\gamma_\mathrm{e}-1}{\gamma_\mathrm{e}+1}\right)^{\gamma}
k_2^2x_0^{-2\lambda-1}\\
&&\times\left\{\frac{\alpha-2\lambda}
{\left[S-(1+\lambda)\eta S'\right]^{\gamma}}
-\gamma(1+\lambda)\eta
\frac{\lambda S'+(1+\lambda)\eta S''}
{\left[S-(1+\lambda)\eta S'\right]^{\gamma+1}}
\right\}.\nonumber
\end{eqnarray}
Therefore, the equation of motion is rewritten in terms of $S(\eta)$ as 
\begin{equation}
(1+\lambda)^2S''=-\frac{2}{\gamma_\mathrm{e}+1}
\left(\frac{\gamma_\mathrm{e}-1}{\gamma_\mathrm{e}+1}\right)^{\gamma}\nonumber
\left\{\frac{\alpha-2\lambda}
{\left[S-(1+\lambda)\eta S'\right]^{\gamma}}
-\gamma(1+\lambda)\eta
\frac{\lambda S'+(1+\lambda)\eta S''}
{\left[S-(1+\lambda)\eta S'\right]^{\gamma+1}}
\right\},
\end{equation}
and we obtain the following ordinary differential equation;
\begin{equation}
S''=\frac{2(\gamma_\mathrm{e}-1)^\gamma}{(1+\lambda)^2}
\frac{(\alpha-2\lambda)[S-(1+\lambda)\eta S']-\gamma\lambda(1+\lambda)\eta S'}
{2\gamma\left(\gamma_\mathrm{e}-1\right)^{\gamma}\eta^2
-(\gamma_\mathrm{e}+1)^{\gamma+1}[S-(1+\lambda)\eta S']^{\gamma+1}},
\label{ode}
\end{equation}
The distribution of physical variables are obtained by solving this equation under
the initial conditions;
\begin{equation}
S(1)=1,\ \ \ S'(1)=\frac{2}{(1+\lambda)(\gamma_\mathrm{e}+1)}.
\label{ini}
\end{equation}

\subsection{Determination of eigen values}
Equation (\ref{ode}) can be integrated to $\eta\rightarrow -\infty$ only for the eigen value 
$\lambda$, 
because the derivative $S''$ diverges at a singular point 
where its denominator becomes zero. 
In other words, the eigen value $\lambda$ is determined so that 
its numerator and denominator of the right hand side of (\ref{ode}) vanish at the singular point. 
Setting the denominator of the left hand side of (\ref{ode}) to be zero, it is found that the value of $\eta$ 
at the singular point is determined by the relation
\begin{equation}
\eta=\left[\frac{(\gamma_\mathrm{e}+1)^{\gamma+1}}{2\gamma(\gamma_\mathrm{e}-1)^\gamma}\right]^{1/2}[S-(1+\lambda)\eta S']^{(\gamma+1)/2}.
\end{equation}
Therefore $\eta$ takes positive value at the singular point. 
We determine the eigen value $\lambda$ as follows. 

Introducing new variables $R$ and $T$ as 
\begin{eqnarray}
\eta^{2/(1+\gamma)}R&=&S,
\label{varR}\\
\eta^{2/(1+\gamma)}T&=&S-(1+\lambda)\eta S',
\label{varT}
\end{eqnarray}
equation (\ref{ode}) is converted to two first order differential equations by the following steps. 
First $S'$ is rewritten with $R$ and $T$ as
\begin{equation}
S'=\frac{1}{1+\lambda}\eta^{(1-\gamma)/(1+\gamma)}(R-T).
\label{S'}
\end{equation}
Then its derivative is
\begin{equation}
S''=\frac{1}{1+\lambda}\eta^{-2\gamma/(1+\gamma)}
\left[\frac{1-\gamma}{1+\gamma}(R-T)+\frac{dR}{d\ln\eta}-
\frac{dT}{d\ln\eta}\right].
\label{S''}
\end{equation}
Substitution of equations (\ref{varT})-(\ref{S''}) into equation 
(\ref{ode}) yields 
\begin{equation}
(1+\lambda)\left[\frac{1-\gamma}{1+\gamma}(R-T)+
\frac{dR}{d\ln\eta}-\frac{dT}{d\ln\eta}\right]=
2\left(\gamma_\mathrm{e}-1\right)^{\gamma}
\frac{[\alpha+(\gamma-2)\lambda]T-\gamma\lambda R}
{2\gamma\left(\gamma_\mathrm{e}-1\right)^{\gamma}
-[(\gamma_\mathrm{e}+1)T]^{\gamma+1}}.
\label{RT}
\end{equation}
On the other hand, the derivative of equation (\ref{varR}) is
\begin{equation}
S'=\eta^{(1-\gamma)/(1+\gamma)}\left(\frac{2}{1+\gamma}R
+\frac{dR}{d\ln\eta}\right).
\end{equation}
Combining this expression with equation (\ref{S'}), we obtain
\begin{equation}
(1+\lambda)\frac{dR}{d\ln\eta}=\left[1-\frac{2(1+\lambda)}{1+\gamma}\right]R-T,
\label{dR}
\end{equation}
and substitution of this equation into equation (\ref{RT}) yields
\begin{equation}
(1+\lambda)\frac{dT}{d\ln\eta}=-\lambda R-\frac{2+(1-\gamma)\lambda}{1+\gamma}T
-2\left(\gamma_\mathrm{e}-1\right)^{\gamma}
\frac{[\alpha+(\gamma-2)\lambda]T-\gamma\lambda R}
{2\gamma\left(\gamma_\mathrm{e}-1\right)^{\gamma}
-[(\gamma_\mathrm{e}+1)T]^{\gamma+1}}.
\label{dT}
\end{equation}
Dividing equation (\ref{dT}) by equation (\ref{dR}), 
we obtain the following ordinary differential equation;
\begin{eqnarray}
\frac{dT}{dR}&=&-
\left\{\lambda R+\frac{2+(1-\gamma)\lambda}{1+\gamma}T+
2\left(\gamma_\mathrm{e}-1\right)^{\gamma}
\frac{[\alpha+(\gamma-2)\lambda]T-\gamma\lambda R}
{2\gamma\left(\gamma_\mathrm{e}-1\right)^{\gamma}
-[(\gamma_\mathrm{e}+1)T]^{\gamma+1}}\right\}
\label{diff_eq}\\
&&\times\left\{\left[1-\frac{2(1+\lambda)}{1+\gamma}\right]R-T\right\}^{-1}.
\nonumber
\end{eqnarray}
The initial conditions for the variables $R$ and $T$ are derived 
from its definition (\ref{varR}),(\ref{varT}), and the condition (\ref{ini}) as
\begin{equation}
R_\mathrm{f}=1,\ \ \ T_\mathrm{f}=\frac{\gamma_\mathrm{e}-1}{\gamma_\mathrm{e}+1}.
\label{front}
\end{equation}
For an arbitrary $\lambda$, the integration of equation (\ref{diff_eq}) from the values of the shock front $(R_
\mathrm{f},T_\mathrm{f})$ can not necessarily reach the other boundary, 
because the derivative $dT/dR$ diverges .
To avoid this situation, we require that the denominator and the numerator of equation (\ref{diff_eq}) 
vanish at the same point, which leads to the following conditions 
\begin{equation}
[\alpha+(\gamma-2)\lambda]T-\gamma\lambda R=0,
\end{equation}
\begin{equation}
2\gamma\left(\frac{\gamma_\mathrm{e}-1}{\gamma_\mathrm{e}+1}\right)^{\gamma}
-(1+\gamma_\mathrm{e})T^{\gamma+1}=0.
\end{equation}
Therefore, we determine the eigen value $\lambda$ so that 
equation (\ref{diff_eq}) can be integrated from $(R_\mathrm{f},T_\mathrm{f})$ to
$(R_\mathrm{s},T_\mathrm{s})$ given by
\begin{equation}
R_\mathrm{s}=\frac{\alpha+(\gamma-2)\lambda}{\gamma\lambda}
\left[\frac{2\gamma}{\gamma_\mathrm{e}+1}\left(\frac{\gamma_\mathrm{e}-1}{\gamma_\mathrm{e}+1}\right)
^{\gamma}\right]^{1/(\gamma+1)},\ \ \ 
T_\mathrm{s}=\left[\frac{2\gamma}{\gamma_\mathrm{e}+1}\left(\frac{\gamma_\mathrm{e}-1}{\gamma_\mathrm{e}+1}\right)
^{\gamma}\right]^{1/(\gamma+1)}.
\label{sing}
\end{equation}

\end{document}